\documentclass{PoS}
\usepackage{amsmath}
\usepackage{graphicx}
\usepackage{amsfonts}
\usepackage{amssymb}

\title{Recent progress on supersymmetric effects\\ in rare K decays}

\ShortTitle{SUSY and rare K decays}

\author{\speaker{Christopher Smith}\thanks{Work supported by the Schweizerischer Nationalfonds.}\\
	  Institut f\"{u}r Theoretische Physik, Universit\"{a}t Bern, CH-3012 Bern, Switzerland\\
	  E-mail: \email{chsmith@itp.unibe.ch}}

\abstract{The dominant MSSM effects in the rare $K$ decays $K^{+}\rightarrow\pi^{+}%
\nu\bar{\nu}$, $K_{L}\rightarrow\pi^{0}\nu\bar{\nu}$, $K_{L}\rightarrow\pi^{0}e^{+}e^{-}$ 
and $K_{L}\rightarrow\pi^{0}\mu^{+}\mu^{-}$, are discussed both within and without 
the minimal flavor violation hypothesis, at moderate and large $\tan\beta$. In each 
case, the sensitivities to MSSM soft-breaking terms are compared, laying emphasis 
on possible correlations among observables. In most scenarios, rare $K$ decays offer
unique windows into the $\Delta S=1$ sector of the soft-breaking terms. Therefore,
together with B-physics and collider observables, these modes will be essential 
for reconstructing the still elusive SUSY-breaking mechanism.}

\FullConference{KAON International Conference\\
 May 21-25 2007\\
 Laboratori Nazionali di Frascati dell'INFN, Rome, Italy}

\begin{document}

\section{Introduction}

The FCNC-induced decays, $K^{+}\rightarrow\pi^{+}\nu\bar{\nu}$, $K_{L}%
\rightarrow\pi^{0}\nu\bar{\nu}$, $K_{L}\rightarrow\pi^{0}e^{+}e^{-}$ and
$K_{L}\rightarrow\pi^{0}\mu^{+}\mu^{-}$, are very suppressed in the Standard
Model (SM), where they can be predicted very accurately\cite{HaischHere}.
Therefore, these modes are ideal for probing possible New Physics
effects\cite{TarantinoHere}. In the present talk, the signatures of
supersymmetry (SUSY), in its simplest realization as the MSSM, are reviewed.
As is well-known, SUSY unifies matter (fermions) and interactions (bosons),
and has a number of desirable features, e.g. it provides a dark matter
candidate, helps unify the gauge couplings at high-energy and stabilizes the electroweak
scale\cite{Martin}. Even though the minimal supersymmetrization of the SM requires one
super-partner for each SM particle (and two Higgs doublets), it is very
constrained and involves only a few free parameters. The problem, however, is
that SUSY must be broken, and the precise mechanism still eludes us.
Therefore, in practice, an effective description is adopted, introducing all
possible explicit soft-breaking terms allowed by the gauge symmetries. In the
squark sector, there are $LL$ and $RR$ mass-terms and trilinear couplings
giving rise to $LR$ mass-terms after the Higgses acquire their VEV's, $\langle
H_{u,d}^{0}\rangle=v_{u,d}$:
\[
\mathcal{L}_{soft}^{LL,RR}=-\tilde{Q}^{\dagger}\mathbf{m}_{Q}^{2}\tilde
{Q}-\tilde{U}\mathbf{m}_{U}^{2}\tilde{U}^{\dagger}-\tilde{D}\mathbf{m}_{D}%
^{2}\tilde{D}^{\dagger},\;\;\mathcal{L}_{soft}^{LR}=-\tilde{U}\mathbf{A}%
^{U}\tilde{Q}H_{u}+\tilde{D}\mathbf{A}^{D}\tilde{Q}H_{d}\;,
\]
with $\tilde{Q}=(\tilde{u}_{L},\tilde{d}_{L})^{T}$, $\tilde{U}=\tilde{u}_{R}%
^{\dagger}$, $\tilde{D}=\tilde{d}_{R}^{\dagger}$. Obviously, $\mathbf{m}%
_{Q,U,D}^{2}$ and $\mathbf{A}^{U,D}$, which are $\mathbf{3\times3}$ matrices
in flavor-space, generate a very rich flavor-breaking sector as squark mass
eigenstates can differ substantially from their gauge eigenstates.\vspace{0.2cm}

\textbf{What to expect from SUSY in rare $K$ decays:} In the SM, the
$Z$-penguin is the dominant contribution, and is tuned by $\lambda_{t}%
=V_{ts}^{\ast}V_{td}$ (Fig.1$a$). The four MSSM corrections depicted in Figs.1$b-e$
(together with box diagrams), represent the dominant corrections, and are thus
the only MSSM effects for which rare $K$ decays can be sensitive probes. Let us
briefly describe each of them. First, there is the charged Higgs contribution to the
$Z$-penguin (Fig.1$b$), which is, at moderate $\tan\beta=v_{u}/v_{d}$, aligned with
the SM one ($\sim\lambda_{t}$). Then, there is the supersymmetrized version of Figs.1$a-b$,
with charginos -- up-squarks in place of $W^{\pm}/H^{\pm}$ -- up-quarks in
the loop (Fig.1$c$), and which is sensitive to the mixings among the six up-squarks
($Z^{U}$), a priori not aligned with the CKM mixings. Another purely
supersymmetric contribution, relevant only for charged lepton modes, is the gluino
electromagnetic penguin (Fig.1$d$), sensitive to down-squark mixings ($Z^{D}$).
The last class of effects consists of neutral Higgs FCNC (Fig.1$e$), and arises at large
$\tan\beta\approx m_{t}/m_{b}\approx50$. Indeed, the 2HDM-II structure of the
Higgs couplings to quarks, required by SUSY, is not preserved beyond leading order due to
$\mathcal{L}_{soft}$, and the ``wrong Higgs'', $H_{u}$, gets coupled to
down-type quarks, $\mathcal{L}_{eff}\supset\bar{d}_{R}^{i}Y_{d}^{ik}(H_{d}%
^{0}+\epsilon Y_{u}^{\dagger}Y_{u}H_{u}^{0\dagger})^{kj}d_{L}^{j}$. Clearly,
once the Higgses acquire their VEV's, there is a mismatch between quark mass
eigenstates and Higgs couplings; both are no longer diagonalized
simultaneously and Higgs FCNC are generated\cite{BabuKolda99}.\vspace{0.2cm}

\textbf{Bottom-up approach and Minimal Flavor Violation: }There are obviously
too many parameters in $\mathcal{L}_{soft}$ to have any hope to fix them all
from rare $K$ decays. At the same time, however, observed FCNC
transitions and CP-violation seem to indicate that new physics induces only
small departures with respect to the SM. Therefore, one starts from a
lowest-order basis in which the flavor-breakings due to $\mathbf{m}%
_{Q,U,D}^{2}$ and $\mathbf{A}^{U,D}$ are minimal. This can take the form of
$mSUGRA$, alignment of squarks with quarks or the Minimal Flavor Violation
hypothesis (MFV). In a second stage, one probes the possible signatures of
departures from this minimal setting. The goal being, ultimately, to constrain
SUSY-breaking models, which imply specific soft-breaking structures. At that
stage, information from rare $K$ decays, colliders and $B$-physics must of
course be combined.

Here we adopt MFV as the lowest order basis, i.e. we impose that the SM
Yukawas $\mathbf{Y}_{u,d}$ are the only sources of flavor-breaking\cite{MFV}.
In practice, this means that $\mathcal{L}_{soft}$ terms can be expanded as
($a_{i},b_{i}\sim O(1)$, $A_{0}$ and $m_{0}$ setting the overall mass-scale
as in $mSUGRA$)%
\begin{align*}
\mathbf{m}_{Q}^{2}  &  =m_{0}^{2}(a_{1}\mathbf{1}+b_{1}\mathbf{Y}_{u}%
^{\dagger}\mathbf{Y}_{u}+b_{2}\mathbf{Y}_{d}^{\dagger}\mathbf{Y}_{d}%
+b_{3}(\mathbf{Y}_{d}^{\dagger}\mathbf{Y}_{d}\mathbf{Y}_{u}^{\dagger
}\mathbf{Y}_{u}+\mathbf{Y}_{u}^{\dagger}\mathbf{Y}_{u}\mathbf{Y}_{d}^{\dagger
}\mathbf{Y}_{d})),\mathbf{m}_{U}^{2}=m_{0}^{2}(a_{2}\mathbf{1}+b_{4}%
\mathbf{Y}_{u}\mathbf{Y}_{u}^{\dagger}),\\
\mathbf{m}_{D}^{2}  &  =m_{0}^{2}(a_{3}\mathbf{1}+b_{5}\mathbf{Y}%
_{d}\mathbf{Y}_{d}^{\dagger}),\mathbf{A}^{U}=A_{0}\mathbf{Y}_{u}%
(a_{4}\mathbf{1}+b_{6}\mathbf{Y}_{d}^{\dagger}\mathbf{Y}_{d}),\mathbf{A}%
^{D}=A_{0}\mathbf{Y}_{d}(a_{5}\mathbf{1}+b_{7}\mathbf{Y}_{u}^{\dagger
}\mathbf{Y}_{u})\;,
\end{align*}
such that all FCNC's and CP-violation are still essentially tuned by the CKM
matrix. For example, the dominant contributions to the $Z$-penguin are those
breaking the $SU(2)_{L}$ gauge-symmetry\cite{NirWorah98,BurasRS98}. In the SM,
this breaking is achieved through a double top-quark mass insertion (Fig.1$f$).
Similarly, in the MSSM, it is the double $\tilde{t}_{L}-\tilde{t}_{R}$ mixing
via the $\mathbf{A}^{U}$ trilinear terms which plays the dominant role (Fig.1$g$ in the
sCKM basis)\cite{ColangeloI98}. Within MFV, this gives a factor $m_{t}%
^{2}\lambda_{t}\left|  a_{4}-\cot\beta\mu^{\ast}\right|  ^{2}$
\cite{IsidoriMPST06}, still enhanced by $m_{t}^{2}$ and tuned by $\lambda_{t}
$.

\begin{figure}[t]
\centering    \includegraphics[width=0.98\textwidth]{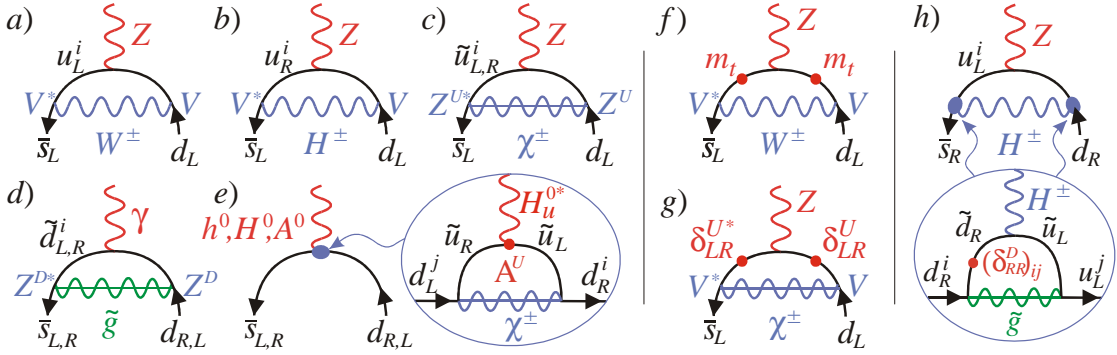}
\caption{$a-e)$ Dominant MSSM contributions to rare $K$ decays. $f-g)$
Dominant sources of $SU(2)_{L}$-breaking in the $Z$-penguin. 
$h)$ Schematic representation of the $H^{\pm}$ contribution to the
$Z$-penguin at large $\tan\beta$.}%
\label{fig1}%
\end{figure}

\section{Supersymmetric effects in $K\rightarrow\pi\nu\bar{\nu}$}

\textbf{SUSY effects in the SM-like operators}, $(\bar{s}d)_{V\pm A}(\bar{\nu
}\nu)_{V-A}$, cannot be distinguished since only $(\bar{s}d)_{V}(\bar{\nu}%
\nu)_{V-A}$ contributes to the $K\rightarrow\pi\nu\bar{\nu}$ matrix-element.
All MSSM effects are thus encoded into a single complex number,
$X^{\nu}\equiv y_{L}^{\nu}+y_{R}^{\nu}$ \cite{BurasRS98}:%
\[
\mathcal{H}_{eff}=y_{L}^{\nu}\left(  \bar{s}d\right)  _{V-A}\left(  \bar{\nu
}\nu\right)  _{V-A}+y_{R}^{\nu}\left(  \bar{s}d\right)  _{V+A}\left(  \bar
{\nu}\nu\right)  _{V-A}\rightarrow\left(  y_{L}^{\nu}+y_{R}^{\nu}\right)
\left(  \bar{s}d\right)  _{V}\left(  \bar{\nu}\nu\right)  _{V-A}\;.
\]

At moderate $\tan\beta$, the dominant MSSM contribution comes from chargino
penguins because of their quadratic sensitivity to up-squark mass-insertions
(Figs.1$c$,1$g$). Within MFV, this means, given the $m_{t}$ enhancement present in
the $\delta_{LR}^{U}$ sector, that $K\rightarrow\pi\nu\bar{\nu}$ are
particularly sensitive. Still, a significant enhancement would require a very
light stop and chargino\cite{IsidoriMPST06}, mostly because of the constraint
from $\Delta\rho$\cite{BurasGGJS00}. Any enhancement $\gtrsim5\%$ would thus
falsify MFV if sparticles are found above $\sim200GeV$, and if $\tan\beta\gtrsim5$
(to get rid of the $H^{\pm}$ contribution). Turning on generic $A^{U}$ terms,
the largest deviations arise in $K\rightarrow\pi\nu\bar{\nu}$, see
Fig.2$a$\cite{IsidoriMPST06}. Further, the decoupling is slower than for
observables sensitive to chargino boxes like $\varepsilon_{K}$. All in all,
given that $K^{+}\rightarrow\pi^{+}\nu\bar{\nu}$ has already been seen, how
large the effect could be for $K_{L}\rightarrow\pi^{0}\nu\bar{\nu}$? By an
extensive, adaptive scanning over the MSSM parameter space, it has been
shown\cite{BurasEJR05} that it is possible to saturate the GN
model-independent bound\cite{GrossmanN97}, which represents a factor
$\sim30$ enhancement of $\mathcal{B}(K_{L}\rightarrow\pi^{0}\nu\bar{\nu})$
over the SM.

At large $\tan\beta$, the chargino contributions decouple, both within and
without MFV, while the Higgs FCNC obviously does not contribute (Fig.1$e$).
However, higher order effects in the $H^{\pm}$ contribution to the $Z$-penguin (Fig.1$h$),
sensitive to $\delta_{RR}^{D}$, can become sizeable beyond MFV\cite{isidoriP06}.
Further, this contribution is slowly decoupling as $M_{H}$ increases compared
to tree-level neutral Higgs exchanges, as for example in $B_{s,d}\rightarrow\mu^{+}\mu^{-}$.
\vspace{0.2cm}

\textbf{SUSY effects in other dimension-six operators}, $(\bar{s}d)(\bar{\nu
}(\mathbf{1},\gamma_{5})\nu)$ and $(\bar{s}\sigma_{\mu\nu}d)(\bar{\nu}%
\sigma^{\mu\nu}(\mathbf{1},\gamma_{5})\nu)$, require active right-handed
neutrinos and will not be discussed here\cite{OtherOperators}. Another
possible class of operators, since the neutrino flavors are not detected, are
$(\bar{s}\Gamma^{A}d)(\bar{\nu}^{i}\Gamma^{B}\nu^{j})$ with $i\neq j$ and
$\Gamma^{A,B}$ some Dirac structures. In the MSSM, such lepton flavor
violating operators arise only from suppressed box diagrams, and cannot lead
to significant effects\cite{GrossmanIM04}. However, they could be sizeable in
the presence of R-parity violating terms\cite{GrossmanIM04,Rparity1}.

\section{Supersymmetric effects in $K_{L}\rightarrow\pi^{0}\ell^{+}\ell^{-}$}

Though the SM predictions for these modes are less accurate than for
$K\rightarrow\pi\nu\bar{\nu}$, they are sensitive to more types of New Physics
operators\cite{MesciaST}. Indeed, the final-state leptons are now charged and
massive. Therefore, besides electromagnetic effects, common to both the muon
and electron modes, the relatively large muon mass opens the possibility to
probe a whole class of helicity-suppressed effects.

\begin{figure}[t]
\centering    \includegraphics[width=0.98\textwidth]{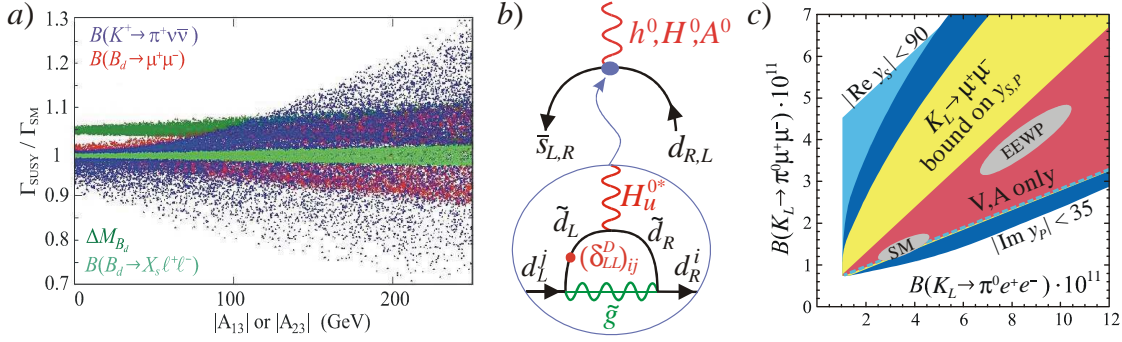}  \vspace
{-0.24cm}\caption{$a)$ Sensitivity of $K^{+}\rightarrow\pi^{+}\nu\bar{\nu}$ to
$\mathbf{A}^{U}$ terms, compared to $B$-physics observables. $b)$ Schematic
representation of the neutral Higgs FCNC beyond MFV, at large $\tan\beta$.
$c)$ Impacts of dim-6 FCNC operators in the
$\mathcal{B}(K_{L}\rightarrow\pi^{0}\mu^{+}\mu^{-})$ vs. 
$\mathcal{B}(K_{L}\rightarrow\pi^{0}e^{+}e^{-})$ plane.}%
\label{fig2}%
\end{figure}\vspace{0.2cm}

\textbf{SUSY effects in the QCD operators}, i.e. in the chromomagnetic
$\bar{s}\sigma_{\mu\nu}dG^{\mu\nu}$ or four-quark operators, have no direct
impact on $K_{L}\rightarrow\pi^{0}\ell^{+}\ell^{-}$.
Indeed, the two-photon CP-conserving contribution is fixed entirely in terms
of the measured $K\rightarrow\pi\pi\pi$, $\pi\gamma\gamma$ modes\cite{LDkpill},
while the indirect CP-violating contribution is fixed from the measured $\varepsilon
_{K}$ and $\mathcal{B}\left(  K_{S}\rightarrow\pi^{0}\ell^{+}\ell^{-}\right)
$\cite{DambrosioEIP98}. At the low-energy scale $(\mu\lesssim m_c)$, new physics
can thus explicitly enter through semi-leptonic FCNC operators only.\vspace{0.2cm}

\begin{table}[t]
\centering    {
\begin{tabular}
[c]{|c|c|c|}\hline
MSSM scenario & $K\rightarrow\pi\nu\bar{\nu}$ & $K_{L}\rightarrow\pi^{0}\ell^{+}\ell^{-}%
$\\\hline
MFV, $\tan\beta\approx2$ & Best sensitivity, but max. & Less sensitive, but
precisely\\
& enhancement $<$ 20-25\% & correlated with $K\rightarrow\pi\nu\bar{\nu}%
$\\\hline
MFV, $\tan\beta\approx50$ & \multicolumn{2}{|c|}{Negligible effects}\\\hline
General, $\tan\beta\approx2$ & Best probes of $\delta_{LR}^{U}$ &
\multicolumn{1}{|l|}{$\delta_{LR}^{U}:$ correlated with $K\rightarrow
\pi\nu\bar{\nu}$}\\
& (quadratic dependence in $\delta_{LR}^{U}$) &
\multicolumn{1}{|l|}{$\delta_{LR}^{D}:$ correlated with $\varepsilon^{\prime
}/\varepsilon$ (but cleaner)}\\\hline
General, $\tan\beta\approx50$ & Good probes of $\delta_{RR}^{D}$ & Good probes
of $\delta_{RR,LL}^{D}$,\\
& (slow decoupling as $M_H \rightarrow \infty$) & corr. with $K_{L}\rightarrow\mu^{+}\mu^{-}$
(but cleaner)\\\hline
\end{tabular}
 }\caption{Sensitivity of rare $K$ decays to MSSM effects, within and without the
MFV hypothesis, and with moderate and large $\tan\beta$. The dominant contributions
can come from single, $(\delta_{j}^{i})_{12}$, and/or double
(e.g. $(\delta_{j}^{i})_{32}^{\ast}(\delta_{j}^{i})_{31}$) mass insertions, see
text for the precise dependences.}%
\label{TableCCL}%
\end{table}

\textbf{SUSY effects in the SM operators}, which are the vector and
axial-vector operators%
\[
\mathcal{H}_{eff}=y_{7V}\left(  \bar{s}d\right)  _{V}\left(  \bar{\ell}%
\ell\right)  _{V}+y_{7A}\left(  \bar{s}d\right)  _{V}\left(  \bar{\ell}%
\ell\right)  _{A}\;,
\]
can in principle be disentangled thanks to the different sensitivities of the
two modes to the axial-vector current (it also produces $\ell^{+}\ell^{-} $ in
a helicity-suppressed $0^{-+}$ state). Various MSSM contributions can enter in
$y_{7A}$ and $y_{7V}$. First, chargino contributions to the $Z$-penguin
(Fig.1$c$) enter as $y_{7A},y_{7V}\sim(\delta_{RL}^{U})_{32}^{\ast}(\delta
_{RL}^{U})_{31}$, and are thus directly correlated to the corresponding
contribution to $K\rightarrow\pi\nu\bar{\nu}$ discussed
previously\cite{IsidoriMPST06,ChoMW96}. Within MFV, the maximal effect for
$K_{L}\rightarrow\pi^{0}\ell^{+}\ell^{-}$ is about one third of the one for
$K_{L}\rightarrow\pi^{0}\nu\bar{\nu}$, hence may be inaccessible due to
theoretical uncertainties. Secondly, gluino contributions to the electromagnetic
operator $\bar{s}\sigma_{\mu\nu}dF^{\mu\nu}$ (Fig.1$d$) can be absorbed into
$y_{7V}\sim(\delta_{RL}^{D})_{12}$. Even if directly correlated with
$\varepsilon^{\prime}/\varepsilon$, sizeable effects in
$K_{L}\rightarrow\pi^{0}\ell^{+}\ell^{-}$ are still possible\cite{BurasCIRS99}.
Finally, $H^{\pm}$ contributions arise at large $\tan\beta$ (Fig.1$h$), with
$y_{7A},y_{7V}\sim(\delta_{RR}^{D})_{12}$, and are directly correlated with
those for $K\rightarrow\pi\nu\bar{\nu}$\cite{isidoriP06}.\vspace{0.2cm}

\textbf{SUSY effects in the (pseudo-)scalar operators}, which can be
helicity-suppressed (i.e., $y\sim m_{\ell}$) or not:%
\[
\mathcal{H}_{eff}=y_{S}\left(  \bar{s}d\right)  \left(  \bar{\ell}\ell\right)
+y_{P}\left(  \bar{s}d\right)  \left(  \bar{\ell}\gamma_{5}\ell\right)
+y_{S}^{\prime}\left(  \bar{s}\gamma_{5}d\right)  \left(  \bar{\ell}%
\ell\right)  +y_{P}^{\prime}\left(  \bar{s}\gamma_{5}d\right)  \left(
\bar{\ell}\gamma_{5}\ell\right)  \;.
\]
The first (last) two operators contribute to $K_{L}\rightarrow\pi^{0}\ell^{+}\ell^{-}$
($K_{L}\rightarrow\ell^{+}\ell^{-}$). In the MSSM at large $\tan\beta$, they
arise from Higgs FCNC\cite{IsidoriRetico}, and are thus helicity-suppressed
(Fig.2$b$). Sizeable effects for the muon mode are possible beyond MFV, where they
are sensitive to $(\delta_{RR,LL}^{D})_{12}$ and
$(\delta_{RR}^{D})_{23}(\delta_{LL}^{D})_{31}$ mass-insertions. Also, even if
this contribution is correlated to the one for $K_{L}\rightarrow\mu^{+}\mu
^{-}$, given the large theoretical uncertainties for this mode, a factor
$\sim4$ enhancement is still allowed (Fig.2$c$)\cite{MesciaST}. On the other hand,
helicity-allowed contributions to these operators do not arise in the MSSM, but
could be generated from R-parity violating couplings. Still, a precise fine-tuning
of these couplings would be needed to have an effect for $K_{L}\rightarrow\pi
^{0}\ell^{+}\ell^{-}$ without overproducing $\mathcal{B}^{\exp}(K_{L}%
\rightarrow e^{+}e^{-})=9_{-4}^{+6}\cdot10^{-12}$\cite{MesciaST}%
.\vspace{0.2cm}

\textbf{SUSY effects in the (pseudo-)tensor operators}, $(\bar{s}\sigma
_{\mu\nu}d)(\bar{\ell}\sigma^{\mu\nu}(\mathbf{1},\gamma_{5})\ell)$, the last possible
dimension-six semi-leptonic FCNC operators, are helicity-suppressed in the MSSM\cite{BobethBKU02}
and, being also phase-space suppressed, do not lead to any significant effect\cite{MesciaST}.
Further, they cannot arise from $R$-parity violating couplings.

\section{Conclusion}

The four rare $K$ decay modes, $K^{+}\rightarrow\pi^{+}\nu\bar{\nu}$,
$K_{L}\rightarrow\pi^{0}\nu\bar{\nu}$, $K_{L}\rightarrow\pi^{0}e^{+}e^{-}$
and $K_{L}\rightarrow\pi^{0}\mu^{+}\mu^{-}$, are the only theoretically clean
windows into the $\Delta S=1$ sector. If SUSY is discovered, the pattern of
deviations they could exhibit with respect to the SM (see Table 1) will be
essential to constrain the MSSM parameter-space, and hopefully unveil the
nature of the SUSY-breaking mechanism.

\end{document}